\documentclass[conference]{IEEEtran}
\IEEEoverridecommandlockouts

\usepackage[
    backend=biber,
    style=ieee,
    citestyle=numeric-comp,
    sorting=none
]{biblatex}

\usepackage[backend=biber,style=ieee,sorting=none]{biblatex}
\usepackage{url}
\usepackage{amsmath,amssymb,amsfonts}
\usepackage{algorithmic}
\usepackage{graphicx}
\usepackage{textcomp}
\usepackage{xcolor}
\usepackage{hyperref}
\def\BibTeX{{\rm B\kern-.05em{\sc i\kern-.025em b}\kern-.08em
    T\kern-.1667em\lower.7ex\hbox{E}\kern-.125emX}}

\begin{document}

\title{Windowed Envelope Statistics for Time-Domain Significant Wave Height Estimation From HF Radar\\
\thanks{The authors would like to acknowledge the support of the Natural Sciences and Engineering Research Council (NSERC) through Discovery Grant number RGPIN-2020-07155 to Dr. Reza Shahidi.}
}

\author{
\IEEEauthorblockN{1\textsuperscript{st} Senal Chandrasekara}
\IEEEauthorblockA{
\textit{Department of Electrical and Computer Engineering} \\
\textit{Memorial University of Newfoundland} \\
St. John's, NL, Canada \\
cchandraseka@mun.ca}
\and
\IEEEauthorblockN{2\textsuperscript{nd} Reza Shahidi}
\IEEEauthorblockA{
\textit{Department of Electrical and Computer Engineering} \\
\textit{Memorial University of Newfoundland} \\
St. John's, NL, Canada \\
rshahidi@mun.ca}
}

\maketitle

\begin{abstract}
Significant wave height (SWH) retrieval from high-frequency (HF) radar
typically relies on a weak second-order Doppler continuum that is
sensitive to noise, interference, and spectral leakage. This letter
presents a Windowed Envelope Statistics Estimator (WESE) that operates
directly on beamformed time-domain voltages. A second-order term obtained
from a Neumann expansion of the rough-surface field equation motivates
quadratic compensation of localized radar features. WESE extracts the
mean, standard deviation, or variance from overlapping windows of the
in-phase, quadrature, or envelope-magnitude sequence, followed by
quadratic compensation, rank ordering, least-squares regression, and
causal smoothing. Evaluation used 335 synchronized hourly observations
from a 13.385-MHz, 12-element WERA system at Argentia, Newfoundland and
Labrador. The optimal configuration used quadrature variance, a 16-sample window, 896 retained chronological samples, and 30-h smoothing, achieving an RMSE of 0.152~m and a Pearson correlation of 0.978. This
represents RMSE reductions of 32.1\% and 18.7\% relative to previously
reported linear and second-order compensated ordered-statistics models,
respectively. The results demonstrate robust time-domain SWH estimation
without explicit Doppler-spectrum construction.
\end{abstract}


\section{Introduction}
\IEEEPARstart{H}{igh-frequency} (HF) ocean radar provides shore-based measurements over extended coastal regions through ground-wave propagation and resonant backscatter from the moving sea surface \cite{PaduanGraber1997,PaduanWashburn2013,BarrickReview1978,Gurgel1999}. Radial-current retrieval uses the strong first-order Bragg lines \cite{Crombie1955,Barrick1972}, whereas conventional Significant Wave Height (SWH) retrieval generally relies on the weaker second-order continuum surrounding them \cite{Barrick1974,Barrick1977Extraction,LipaBarrick1980,LipaBarrick1986}. This continuum is commonly 20--40~dB below the Bragg return, distributed across Doppler frequency, and difficult to separate exactly from first-order leakage, background noise, radio-frequency interference, and ship or ionospheric echoes \cite{HuangGill2010,Wyatt2000,WyattGreenMiddleditch2011}.

Full-continuum methods recover a directional wave spectrum by inverting a nonlinear second-order scattering integral \cite{Hasselmann1971,Lipa1978,Wyatt1986,Wyatt1990}. The problem maps a two-dimensional directional spectrum into a one-dimensional Doppler spectrum and is consequently non-unique and noise sensitive. Reduced approaches instead integrate selected second-order regions or use spectral landmarks \cite{Barrick1977Extraction,LipaBarrick1986,WyattGreen2023}. Both require reliable construction and partitioning of a Doppler
spectrum, which becomes difficult when the second-order return is
diffuse, broadened, or partially or fully submerged beneath the noise
floor.

Direct temporal methods avoid explicit second-order-continuum extraction. Shahidi and Gill \cite{ShahidiGill2020} established that first-order received-field statistics can be related to SWH. Hashemi \emph{et al.} \cite{Hashemi2023} subsequently showed that the windowed variance of the received-field upper envelope contains SWH-dependent information. A linear ordered-statistics estimator then mapped high-ranked field voltages to buoy SWH by closed-form regression \cite{ChandrasekaraShahidi2025}, and a quadratic extension introduced a nonlinear compensation stage for second-order scattering \cite{ChandrasekaraShahidi2026}.

This letter introduces a Windowed Envelope Statistics Estimator (WESE) by applying localized statistics on the received radar voltages before
quadratic compensation and ordered regression. Its contributions are
linking the second Neumann term to quadratic SWH dependence, comparing
windowed statistics across \(I\), \(Q\), and envelope representations,
and chronological field validation against linear and quadratic
point-sample estimators.

\section{Spectral and Time-Domain Background}
For a random sea, the SWH \(H_{\mathrm{s}}\approx4\sqrt{m_0}\), where \(m_0\)
is the zeroth moment of the elevation spectrum
\cite{Holthuijsen2007,LonguetHiggins1952,LonguetHiggins1957}. In
Barrick's reduced spectral formulation, SWH is estimated from the
weighted second- to first-order spectral-energy ratio as
\cite{Barrick1977Extraction}
\begin{equation}
H_{\mathrm{s}}
\approx
\frac{2}{k_0}
\sqrt{
\frac{
\displaystyle\int_{\Omega_2}
\frac{\sigma(\omega)}
{w(\omega/\omega_B)}\,d\omega
}{
\displaystyle\int_{\Omega_1}
\sigma(\omega)\,d\omega
}
},
\label{eq:barrick}
\end{equation}
where \(k_0\) is the radar electromagnetic wavenumber,
\(\sigma(\omega)\) is the Doppler power spectrum, \(\omega_B\) is the
Bragg angular frequency, \(\Omega_1\) and \(\Omega_2\) are the selected
first- and second-order spectral regions, respectively, and
\(w(\cdot)\) compensates for frequency-dependent coupling. In field
spectra, \(\Omega_2\) is not an independently observed second-order
cross section; it may contain first-order leakage, noise, or
interference, and its boundaries may be ambiguous.

Instead, the direct time-domain formulation begins with the first-order relation
\begin{equation}
E_{0n}^{+(1)}(\rho,\theta,t)
\propto
\xi(\rho,\theta,t),
\label{eq:first_order_time_domain}
\end{equation}
where \(E_{0n}^{+(1)}\) is the first-order normal component of the
surface electric field evaluated from the upper medium,
\((\rho,\theta)\) specifies the range and azimuth of the ocean
scattering point, \(t\) denotes time, and \(\xi\) is the corresponding
sea-surface displacement \cite{ShahidiGill2020}. Under the
Rayleigh-envelope interpretation, the expected \(k\)th ordered voltage
magnitude used by the linear estimator satisfies
\cite{ChandrasekaraShahidi2025}
\begin{equation}
\mathbb{E}\!\left[
E_{0,n,(\mathrm{upper},k,N)}^{+}
\right]
=
\frac{\mu}{K}
\sqrt{
2\ln\!\left(
\frac{1}{1-\sqrt{k/N}}
\right)
},
\label{eq:prior_orderstat}
\end{equation}
where \(E_{0,n,(\mathrm{upper},k,N)}^{+}\) denotes the \(k\)th
ascending order statistic of the upper-envelope voltage magnitudes
formed from \(N\) samples, \(k\in\{1,\ldots,N\}\) is the rank index,
and \(\mathbb{E}[\cdot]\) denotes the expectation operator. The parameter \(\mu\) is
the Rayleigh scale parameter of the sea-surface envelope, while
\(K>0\) is a real system scaling constant relating the surface
displacement to the received voltage. Since
\(\mu\propto H_{\mathrm{s}}\), where \(H_{\mathrm{s}}\) is the
SWH, the expected ordered voltage magnitudes are
also linearly-proportional to \(H_{\mathrm{s}}\). This relationship
motivates linear fusion of the highest-ranked voltage samples and
produced an RMSE of 0.224~m on field data from Argentia, Newfoundland, Canada \cite{ChandrasekaraShahidi2025}. Hereafter, this
implementation is referred to as the Linear Ordered Statistic Estimator
(LOSE).

Hashemi \emph{et al.} \cite{Hashemi2023} incorporated first- and second-order terms through 
\begin{equation}
 E(t)\approx A\xi^2-A+B\xi+C,
 \label{eq:hashemi_field}
\end{equation}
and reported the finite-window upper-envelope relation
\begin{multline}
 \operatorname{Var}\!\left[E_{\mathrm{upper}}(t)\right]\approx
 \frac{16A^2}{H_s^2}+\frac{B^2 H_s^2(4-\pi)}{32}+C\\
 +2AB\!\left(\frac{3\sqrt{\pi}\, H_s^{3/2}}{32}
 -\sqrt{\frac{\pi}{2}}-\frac{\sqrt{2\pi\ H_s}}{4}\right).
 \label{eq:hashemi_variance}
\end{multline}

A floating-platform simulation study reported SWH RMSEs of
0.10--0.14~m without conventional Doppler-spectrum computation
\cite{Hashemi2023}. Chandrasekara and Shahidi subsequently modelled
radar features quadratically and applied the physically admissible inverse
\cite{ChandrasekaraShahidi2026}:
\begin{equation}
E\simeq AH_s^2+BH_s+C,\qquad
\widehat H_s=\frac{-B+\sqrt{B^2+4A(E-C)}}{2A}.
\label{eq:prior_quadratic_model}
\end{equation}
Regression of the rank-ordered estimates yielded a minimum RMSE of
0.187~m.

Building on these
developments, the proposed WESE framework first replaces individual
radar-envelope samples with localized statistical descriptors and
then applies constrained quadratic compensation and rank-ordered
regression. It therefore incorporates localized statistical
aggregation directly into the ordered-statistics estimation pipeline,
rather than using the aggregation only as a separate preprocessing
operation.

\section{Methodology}

The proposed WESE framework first retains \(M\) chronological samples
from each hourly radar-voltage record, partitions them into localized
windows, and extracts a statistical descriptor from each window. This windowing exploits the separation between rapid
sample-to-sample voltage fluctuations and the comparatively slow
evolution of the underlying sea state. Within each short window, the
scattering process is treated as locally quasi-stationary, allowing
multiple voltage samples to be aggregated into a more representative
sea-state feature than an individual sample. Quadratic compensation is
then applied to the resulting descriptors, followed by descending rank
ordering, least-squares regression, and causal smoothing.

\subsection{Localized Statistical Features}

For the $i^{th}$ hourly radar record, let \(x_i[n]\) denote the selected
real-valued representation. Each record contains \(N\) chronological
samples, of which the first \(M\leq N\) are retained. Windows of length
\(L\), advanced by \(D\) samples, produce
\[
R=\left\lfloor\frac{M-L}{D}\right\rfloor+1
\]
localized descriptors. Let \(\mathcal{W}_{i,r}\) denote the $r^{th}$
window, where \(r=0,\ldots,R-1\).

The local mean, windowed variance, and sample standard deviation
are respectively defined as
\begin{align}
\mu_{i,r}
&=
\frac{1}{L}
\sum_{n=0}^{L-1}x_i[rD+n],
\label{eq:mean}\\
\upsilon_{i,r}
&=
\frac{1}{L}
\sum_{n=0}^{L-1}
\left(
x_i[rD+n]-\mu_{i,r}
\right)^2,
\label{eq:variance}\\
\varsigma_{i,r}
&=
\sqrt{\upsilon_{i,r}}.
\label{eq:std}
\end{align}
For descriptor type
\(\alpha\in\{\mathrm{mean},\mathrm{std},\mathrm{var}\}\), the resulting
localized feature is written as
\begin{equation}
s_{i,r}^{(\alpha)}
=
G_{\alpha}\!\left(\mathcal{W}_{i,r}\right),
\label{eq:selected_local_feature}
\end{equation}
where \(G_{\mathrm{mean}}\), \(G_{\mathrm{std}}\), and
\(G_{\mathrm{var}}\) return \(\mu_{i,r}\), \(\varsigma_{i,r}\), and
\(\upsilon_{i,r}\), respectively.

\subsection{Physical Motivation for Quadratic Compensation}

In the Walsh whole-space formulation \cite{Gill1999,WalshGill2000}, the rough air--sea interface is
represented by \(z=\xi(x,y)\), and the two media are embedded within a
single domain through the Heaviside function
\begin{equation}
h\!\left[z-\xi(x,y)\right]
=
\begin{cases}
0, & z\leq \xi(x,y),\\
1, & z> \xi(x,y).
\end{cases}
\label{eq:heaviside}
\end{equation}
The spatially varying conductivity and permittivity are written as
\begin{equation}
\begin{aligned}
\sigma(\mathbf r)
&=
\left[1-h\!\left(z-\xi\right)\right]\sigma_w,\\
\varepsilon(\mathbf r)
&=
\varepsilon_0 h\!\left(z-\xi\right)
+\varepsilon_w\left[1-h\!\left(z-\xi\right)\right],
\end{aligned}
\label{eq:whole_space_material_parameters}
\end{equation}
where \(\sigma_w\) and \(\varepsilon_w\) denote the conductivity and
permittivity of seawater, respectively, and \(\varepsilon_0\) is the
free-space permittivity. With \(\mathbf E^{+}\) and
\(\mathbf E^{-}\) denoting the electric fields above and below the
interface, respectively, the total field is expressed as
\cite{Gill1999,WalshGill2000}
\begin{equation}
\mathbf E
=
h\mathbf E^{+}
+
(1-h)\mathbf E^{-}.
\label{eq:whole_space}
\end{equation}

Substitution of Equation~\ref{eq:whole_space} into Maxwell's equations
transforms the rough-boundary problem into the whole-space source
equation
\begin{equation}
\nabla^2\mathbf E+\gamma_0^2\mathbf E
=
\frac{\eta_r^2-1}{\eta_r^2}
\nabla\!\left[
\left(\mathbf n\cdot\mathbf E^{+}\right)
\delta\!\left(z-\xi\right)
\right]
-\Gamma_s^E(\mathbf J_s),
\label{eq:whole_wave}
\end{equation}
where \(\gamma_0\) is the propagation constant in the upper medium,
\(\eta_r\) is the complex relative refractive index of seawater,
\(\mathbf n\) is the surface-normal vector, \(\delta(\cdot)\) is the
Dirac delta function, \(\mathbf J_s\) is the prescribed source-current
density, and \(\Gamma_s^E\) is the source operator mapping
\(\mathbf J_s\) to the corresponding electric-field forcing term
\cite{Gill1999,WalshGill2000}. Application of the corresponding Green function,
followed by a horizontal Fourier transformation, the good-conductor
approximation, and a small-slope reduction, yields
\begin{equation}
E_{0n}^{+}
-
\mathcal{T}_1\!\left(E_{0n}^{+}\right)
=
E_{\mathrm{s}},
\label{eq:operator}
\end{equation}
where \(E_{0n}^{+}\) denotes the normal component of the surface
electric field evaluated from the upper medium, \(E_{\mathrm{s}}\) is
the source-field contribution, and \(\mathcal{T}_1\) is the
rough-surface interaction operator
\cite{Gill1999}.

Rearranging Equation~\ref{eq:operator} and applying successive substitution
gives the Neumann expansion
\begin{equation}
E_{0n}^{+}
=
\sum_{m=0}^{\infty}
\mathcal{T}_1^{m}\!\left(E_{\mathrm{s}}\right),
\label{eq:neumann}
\end{equation}
where \(\mathcal{T}_1^{0}\) denotes the identity operator. Under the
small-slope approximation, the first- and second-order scattered-field
contributions satisfy
\begin{align}
E_{0n}^{+(1)}
&=
\mathcal{T}_1\!\left(E_{\mathrm{s}}\right)
\propto \xi,
\label{eq:first_order}\\
E_{0n}^{+(2)}
&=
\mathcal{T}_1^{2}\!\left(E_{\mathrm{s}}\right)
=
\mathcal{T}_1\!\left(E_{0n}^{+(1)}\right)
\propto \xi^2.
\label{eq:second_order}
\end{align}
For the second-order approximation
\(E_{0n}^{+}\approx a_0+a_1\xi+a_2\xi^2\), a mean-removed sea satisfies
\(\mathbb{E}[\xi]=0\) and
\(\mathbb{E}[\xi^2]=\sigma_\xi^2\approx H_{\mathrm{s}}^2/16\).
Therefore,
\begin{equation}
\mathbb{E}\!\left[E_{0n}^{+}\right]
\approx a_0+\frac{a_2}{16}H_{\mathrm{s}}^2.
\label{eq:mean_quad}
\end{equation}

Equation~\ref{eq:mean_quad} establishes the physical motivation for
introducing a quadratic sea-state dependence into the empirical
calibration. In this estimator, the localized descriptors in Equation \ref{eq:selected_local_feature} are
formed first, after which each candidate descriptor is calibrated using
\begin{equation}
s_{i,r}^{(\alpha)}
\simeq
A_{\alpha}H_{\mathrm{s},i}^{2}+C_{\alpha}.
\label{eq:descriptor_quadratic_model}
\end{equation}
This descriptor-level relationship is treated as a physically motivated
empirical correction rather than a complete inversion of the
second-order radar cross section.

\subsection{Quadratic Compensation, Rank Regression, and Smoothing}

For a fixed radar representation and descriptor type, the superscript
\(\alpha\) is omitted hereafter for compactness. The constrained model
in Equation~\ref{eq:descriptor_quadratic_model} is fitted using the training
descriptors. With vectorized training descriptors
\(\mathbf{s}_{\mathrm{tr}}\) and the corresponding buoy targets repeated
for each localized window in \(\mathbf{h}_{\mathrm{rep}}\), the fitted
coefficients are
\begin{equation}
\begin{bmatrix}
\widehat A\\
\widehat C
\end{bmatrix}
=
\begin{bmatrix}
\mathbf{h}_{\mathrm{rep}}^{\circ 2} & \mathbf{1}
\end{bmatrix}^{\dagger}
\mathbf{s}_{\mathrm{tr}},
\label{eq:quadfit}
\end{equation}
where \((\cdot)^\dagger\) denotes the Moore--Penrose pseudoinverse and
\((\cdot)^{\circ 2}\) denotes element-wise squaring. Configurations for
which \(\widehat A\leq10^{-12}\) are rejected. Each localized descriptor
is then transformed into the nonnegative compensated feature
\begin{equation}
c_{i,r}
=
\sqrt{
\max\!\left(
\frac{s_{i,r}-\widehat C}{\widehat A},
0
\right)
}.
\label{eq:compensation}
\end{equation}

\begin{figure}[h]
\centering
\includegraphics[width=\columnwidth]{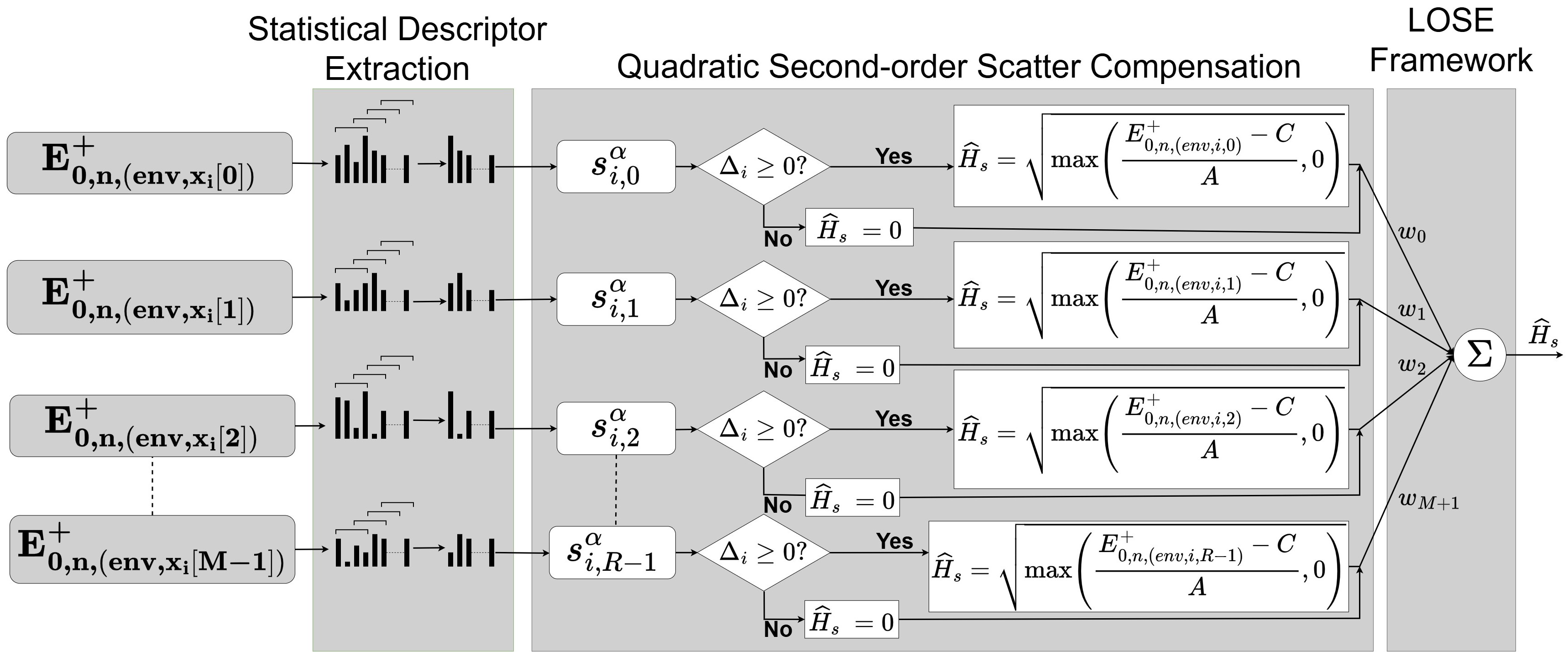}
\caption{Processing architecture of the proposed WESE framework.}
\label{fig:architecture}
\end{figure}

For each hourly record, the \(R\) compensated descriptors are sorted in
descending order. Stacking the resulting feature vectors as the rows of
\(\mathbf Q_{\mathrm{tr}}\) and \(\mathbf Q_{\mathrm{te}}\), the regression
weights and test predictions are obtained as
\begin{equation}
\mathbf w_{\mathrm{opt}}
=
\mathbf Q_{\mathrm{tr}}^{\dagger}\mathbf h_{\mathrm{tr}},
\qquad
\widehat{\mathbf h}_{\mathrm{te}}
=
\mathbf Q_{\mathrm{te}}\mathbf w_{\mathrm{opt}}.
\label{eq:regression}
\end{equation}
Finally, a causal moving average of length \(W\) is applied:
\begin{equation}
\overline H_{\mathrm{s}}[i]
=
\frac{1}{W}
\sum_{p=0}^{W-1}
\widehat H_{\mathrm{s}}[i-p].
\label{eq:smoothing}
\end{equation}
with initial-value padding where preceding estimates are unavailable.
The complete WESE processing architecture is shown in
Fig.~\ref{fig:architecture}.

\section{Experiments}
Measurements were collected from July 5--18, 2018 using a shore-based
monostatic WERA system at Argentia, Newfoundland. The radar
operated at 13.385~MHz and used a 12-element vertically polarized
monopole ULA with \(0.45\lambda_0\) spacing, 40 range gates, and 3~km
range resolution. Bartlett beamforming and range--azimuth interpolation
isolated the Red Island Shoal buoy location. Four 15-minute sequences were
concatenated to produce one hourly complex sequence containing 4096
samples. After removal of one unmatched radar hour, 335 synchronized
radar--buoy pairs remained.

The first \(N_{\mathrm{tr}}=219\) hours were used for calibration and
regression, while the remaining 116 hours formed the chronological
evaluation interval; no evaluation observation was used during fitting.
The search was independently performed for the \(I\), \(Q\), and
\(\lvert I+jQ\rvert\) representations and the local mean, standard
deviation, and variance. Window lengths
\(L\in\{8,16,32,64,128\}\) were evaluated using a fixed step of
\(D=2\) samples. The number \(M\) of retained chronological samples was varied from \(L\)
to the available sequence limit, and the causal smoothing
length was varied over \(W=2,\ldots,30\) hours. Configurations satisfying
\(\widehat A\leq10^{-12}\) were discarded. Performance over the 116-h chronological evaluation interval was
assessed using the RMSE and Pearson correlation
coefficient \(r\).

\section{Results and Discussion}
Table~\ref{tab:results} reports the best configuration for each descriptor--representation pair. Quadrature variance produced the lowest RMSE, using $L=16$, $M=896$, and $W=30$. Mean features from the individual $I$ and $Q$ channels produced no admissible configuration because their fitted coefficient did not satisfy $A>0$; this was a model-compatibility outcome rather than an iterative convergence failure.

\begin{table}[h]
\caption{Minimum-RMSE WESE Configurations}
\label{tab:results}
\centering
\scriptsize
\setlength{\tabcolsep}{3.2pt}
\begin{tabular}{|l|c|r|r|r|c|}\hline
Statistic & Signal & $L$ & $M$ & $W$ & RMSE (m) \\ \hline
Mean & $I$ & -- & -- & -- & -- \\
Mean & $Q$ & -- & -- & -- & -- \\
Mean & $|I+jQ|$ & 32 & 2224 & 30 & 0.157 \\ \hline
Std. dev. & $I$ & 16 & 3158 & 30 & 0.160 \\
Std. dev. & $Q$ & 16 & 804 & 26 & 0.170 \\
Std. dev. & $|I+jQ|$ & 64 & 2196 & 30 & 0.171 \\ \hline
Variance & $I$ & 32 & 1270 & 26 & 0.177 \\
Variance & $Q$ & 16 & 896 & 30 & \textbf{0.152} \\
Variance & $|I+jQ|$ & 64 & 1712 & 30 & 0.183 \\ \hline
\end{tabular}
\end{table}

The optimal prediction tracked the initial reduction in buoy SWH, the
low-energy interval, and the subsequent recovery. As summarized in
Table~\ref{tab:baseline_comparison}, its 0.152~m RMSE is 32.1\% below
the 0.224~m linear ordered-statistics result and 18.7\% below the
0.187~m second-order compensated result. Since all three methods used
the same Argentia campaign and chronological training boundary, these
reductions directly quantify the benefit of replacing isolated point
samples with localized statistical features.

\begin{table}[!t]
\caption{Field-Data Comparison With Prior Ordered-Statistics Models}
\label{tab:baseline_comparison}
\centering
\footnotesize
\begin{tabular}{lcc}
\hline
Method & Feature stage & RMSE (m) \\
\hline
Linear model \cite{ChandrasekaraShahidi2025}
& Point samples & 0.224 \\
Second-order model \cite{ChandrasekaraShahidi2026}
& Compensated point samples & 0.187 \\
Proposed WESE
& Windowed statistics & \textbf{0.152} \\
\hline
\end{tabular}
\end{table}

\begin{figure}[h]
\centering
\includegraphics[width=\columnwidth]{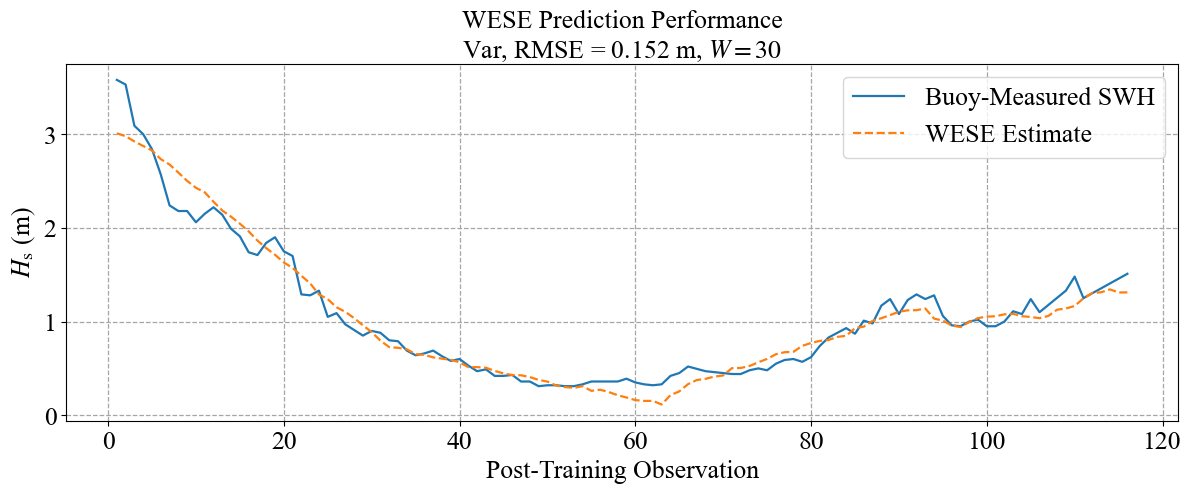}
\vspace{1mm}
\includegraphics[width=\columnwidth]{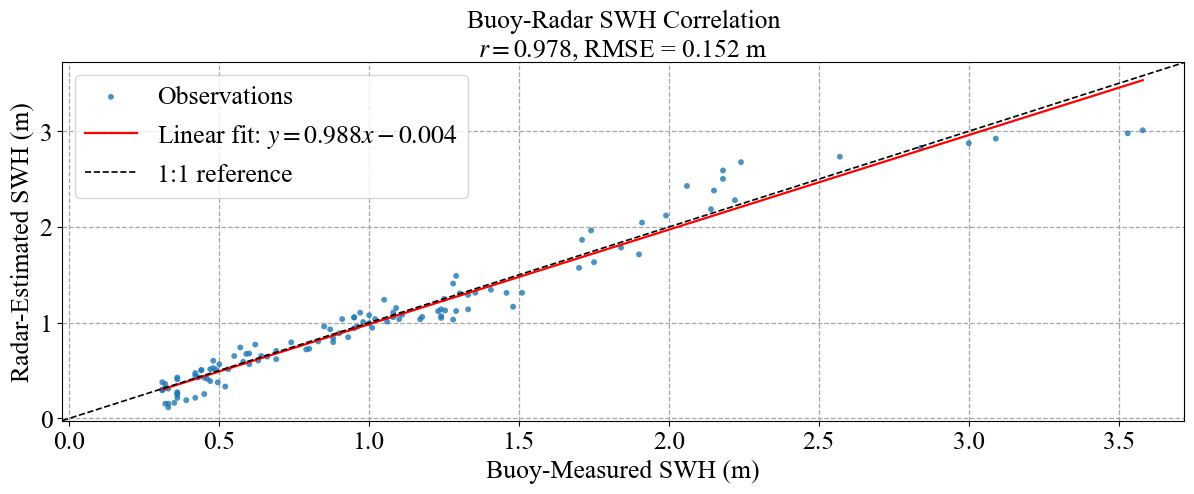}
\caption{Performance of the optimal quadrature-variance WESE
configuration. Top: buoy-measured and radar-estimated SWH over the
evaluation interval. Bottom: radar-estimated versus buoy-measured SWH
with the one-to-one reference line.}
\label{fig:results}
\end{figure}

Figure~\ref{fig:results} shows the buoy--radar SWH comparison over the
evaluation interval and the corresponding predicted-versus-measured
relationship. The Pearson correlation coefficient was \(r=0.978\), and
the fitted relationship between the radar estimate and buoy measurement
was
\begin{equation}
\widehat H_{\mathrm{s}}
=
0.988H_{\mathrm{s}}-0.004,
\label{eq:fit}
\end{equation}
with a mean residual of \(-0.0168\)~m and no sustained long-term drift.

\begin{figure}[!t]
    \centering
    \includegraphics[width=\linewidth]{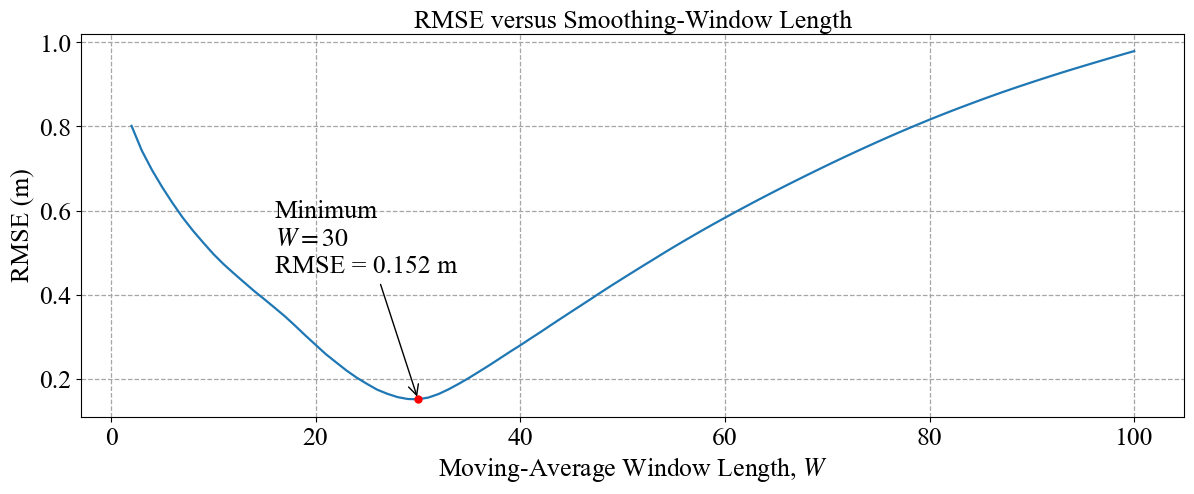}
    \caption{RMSE as a function of causal moving-average window length for the optimal WESE configuration.}
    \label{fig:wese_rmse_window}
\end{figure}

Further examining the temporal averaging revealed that the RMSE initially decreased as $W$ increased, reaching its minimum value at $W=30$, and further increased for longer smoothing windows, as observed in Figure~\ref{fig:wese_rmse_window}.

\section{Conclusion}
WESE provides a direct time-domain SWH estimator that combines overlapping local voltage statistics with physically motivated second-order compensation, ordered features, closed-form regression, and causal smoothing. The Walsh--Neumann expansion explains why the first nonlinear rough-surface correction introduces a squared-elevation term and therefore supports the constrained quadratic mapping used by the estimator. Field experiments on 335 synchronized radar--buoy hours showed that quadrature-component variance achieved 0.152~m RMSE and $r=0.978$. The improvement compared to the previous models in \cite{ChandrasekaraShahidi2025} and \cite{ChandrasekaraShahidi2026} demonstrates that short-time statistical structure contains more robust sea-state information than isolated voltage samples. Future work should test independent sites and broader sea states, separate parameter selection from final evaluation, and investigate regularized regression and reduced-lag smoothing.

\section*{Data Availability}
The buoy significant-wave-height measurements and corresponding
Bartlett-beamformed HF-radar I/Q data used in this study are publicly
available at: \href{https://drive.google.com/drive/folders/1iqM62JDvG1VAjQ5rmmwtZ3gmEvfLGvpj?usp=drive_link}{public dataset repository}.

\printbibliography

\end{document}